\documentclass[11pt]{article}
\usepackage{url}
\usepackage{lineno}
\usepackage{soul}
\usepackage{geometry}
\nolinenumbers
\usepackage{authblk}
\usepackage{hyperref}
\usepackage{graphicx}

\usepackage[font=small, labelfont=bf, skip=6pt]{caption} 

\usepackage{apacite}
\usepackage{regexpatch}
\makeatletter
\xpatchcmd{\@@cite}{\def\BCA##1##2{{\@BAstyle ##1}}}{\def\BCA##1##2{{\@BAstyle ##2}}}{}{}
\makeatother

\hypersetup{
    colorlinks=true,
    linkcolor=black,
    citecolor=black,
    filecolor=black,      
    urlcolor=black,
    pdftitle={Rigorous uncertainty quantification of probabilistic AI weather forecasts with conformal prediction},
    pdfpagemode=FullScreen,
}

\newcommand{\correspondingauthor}{\thanks{Corresponding author. Email: \texttt{aasch@uchicago.edu}}}

\title{\textbf{Rigorous uncertainty quantification of probabilistic AI weather forecasts with conformal prediction}}

\author[1]{\textbf{Anna Asch}\correspondingauthor}
\author[2]{\textbf{Raphael Rossellini}}
\author[1,3]{\textbf{Pedram Hassanzadeh}}
\author[1,2,4]{\textbf{Rebecca Willett}}

\affil[1]{Committee on Computational and Applied Mathematics, University of Chicago}
\affil[2]{Department of Statistics, University of Chicago}
\affil[3]{Department of the Geophysical Sciences, University of Chicago}
\affil[4]{Department of Computer Science, University of Chicago}
\date{\vspace{-6ex}}

\usepackage{amssymb, amsmath}

\begin{document}

\maketitle

\begin{abstract}
Probabilistic weather forecasting is undergoing rapid transformation with artificial intelligence (AI). In traditional numerical weather prediction, computing power can limit how well ensemble forecasts approximate the unknown statistical distribution of future states. AI models facilitate larger ensembles and are trained with probabilistic considerations, ideally leading to better uncertainty quantification. Forecasts from these state-of-the-art models are often considered well-calibrated. However, here we show that the {\it statistical coverage} of such models, the ultimate measure of calibration, can struggle, especially on extreme events. To address this shortcoming, we employ conformal prediction, a class of statistical methods that mathematically guarantees coverage under no distributional assumptions, unlike previous post-processing techniques. We apply online conformal prediction to temperature and precipitation forecasts (including extremes) of three leading global weather models, GenCast, NeuralGCM, and AIFS-ENS, ensuring calibrated uncertainty at no expense to other probabilistic metrics. This post-processing method can be applied to any forecasting model.
\end{abstract}

%
%

\section{Introduction}
Ensemble weather predictions are an essential product of operational forecasting centers. Errors from imperfect observations, data assimilation, and predictive models are amplified by the inherent chaos and multi-scale nature of the global weather system, necessitating forecasts that specify a distribution over future states \cite{lorenz_predictability_1969, palmer_predicting_2000}. But at the heart of probabilistic forecasting is the need for correct uncertainty quantification (UQ). In traditional numerical weather prediction (NWP), empirical distributions are constructed by propagating an ensemble with carefully designed initial condition perturbations through a model \cite{toth_ensemble_1997, molteni_ecmwf_1996, isaksen2010ensemble}. Recent models also incorporate stochastic physics schemes \cite{berner_stochastic_2017}. Still, inevitable initial condition and modeling errors prevent the forecast distribution from matching the true, unknown distribution of future weather states \cite{slingo_uncertainty_2011}. 

A new generation of weather forecasting models, based entirely or partially on AI, improves upon NWP skill at a fraction of the real-time computational cost \cite{pathak_fourcastnet_2022, bi_accurate_2023, lam_learning_2023, kochkov_neural_2024}. Fast inference enables larger ensembles and probabilistic objectives shape the architecture and training, ideally leading to better UQ. However, ensemble creation approaches are largely ad hoc; methods include sampling from a conditional diffusion model \cite{price_probabilistic_2025, couairon_archesweathergen_2026}, evolving flow-dependent perturbations in a learned latent space \cite{chen_machine_2024}, and using traditional methods like bred vectors \cite{bano-medina_toward_2025}. To improve the quality of ensemble generation in AI-based probabilistic forecasting, practitioners have recently begun incorporating the continuous ranked probability score (CRPS) into training objectives \cite{lang_aifs-crps_2026} (probabilistic metrics are defined in Section~S1). The CRPS is strictly proper---it is optimized when the distribution represented by the probabilistic forecaster is the same as the distribution underlying the training data \cite{gneiting2007strictly}. Because of this characterization, CRPS-trained models are generally considered to produce probabilistic forecasts that accurately represent the underlying data distribution, and evaluation studies based on the CRPS, spread-skill ratio (SSR), and rank histogram have supported this interpretation \cite{fortin_why_2014, wilks_reliability_2011}.

In this paper, we examine the calibration of probabilistic weather forecasts. A forecast has correct ``statistical coverage'' if the actual weather falls within its predicted range as often as the model claims it will. For instance, if a model produces $90\%$ temperature prediction intervals, the observed temperature should fall within those bounds $90\%$ of the time. A calibrated model is defined as one that achieves the correct coverage for all forecast intervals simultaneously, i.e., exhibits the correct reliability diagram \cite{hamill_reliability_1997}. Calibration is an important property for socio-economic decision making; for instance, in agricultural applications, falsely confident forecasts of the likelihood of rainfall can cause significant harm \cite{burlig_value_2024, aitken_designing_2026}. Probabilistic models may have good CRPS or SSR values but still fail to be calibrated \cite{hersbach_decomposition_2000}.

To improve calibration, one can post-process forecasts. A classic technique is the ensemble model output statistics (EMOS) method, which quantifies uncertainty via a Gaussian distribution, with parameters depending on the ensemble mean and variance. Variants fit a non-Gaussian distribution \cite{scheuerer_probabilistic_2014}, or learn a nonlinear relationship between the covariates and Gaussian parameters via a parameterized neural network \cite{rasp_neural_2018, hohlein_postprocessing_2024}. To avoid parametric assumptions on the true distribution, several nonparametric methods have been proposed \cite{taillardat_calibrated_2016, bremnes_ensemble_2020, henzi_isotonic_2021, gronquist_deep_2021}. Recent work indicates that post-processing methods, such as model blending, can improve the statistical properties of the ensembles produced by AI weather models \cite{trotta_statistical_2025, hua_improving_2025, aitken_designing_2026}.

None of the aforementioned post-processing methods come with statistical coverage guarantees. Recent advancements in conformal prediction have produced online methods that adapt to distribution shifts and yield \textit{guaranteed} calibration, under no distributional assumptions. To the best of our knowledge, no other study has applied online conformal prediction to ensemble weather forecasts, and, in general, conformal prediction has seen little use in meteorology. Existing work applying conformal prediction to weather forecasts either constructed intervals around point estimates \cite{walz_easy_2024} or worked with fixed adjustments that could not adapt to non-stationarity \cite{gopakumar_uncertainty_2026}. Other use of conformal prediction has thus far been for specific downstream tasks, such as estimating photovoltaic power \cite{renkema_enhancing_2024}, forecasting tropical cyclones \cite{meng_uncertainty_2024,chen_meteorologically-informed_2025}, estimating short-term wind speed \cite{althoff_evaluation_2023}, and sub-grid-scale parameterizations \cite{simm_calibrated_2026}. 

\begin{figure}
\noindent\includegraphics[width=\textwidth]{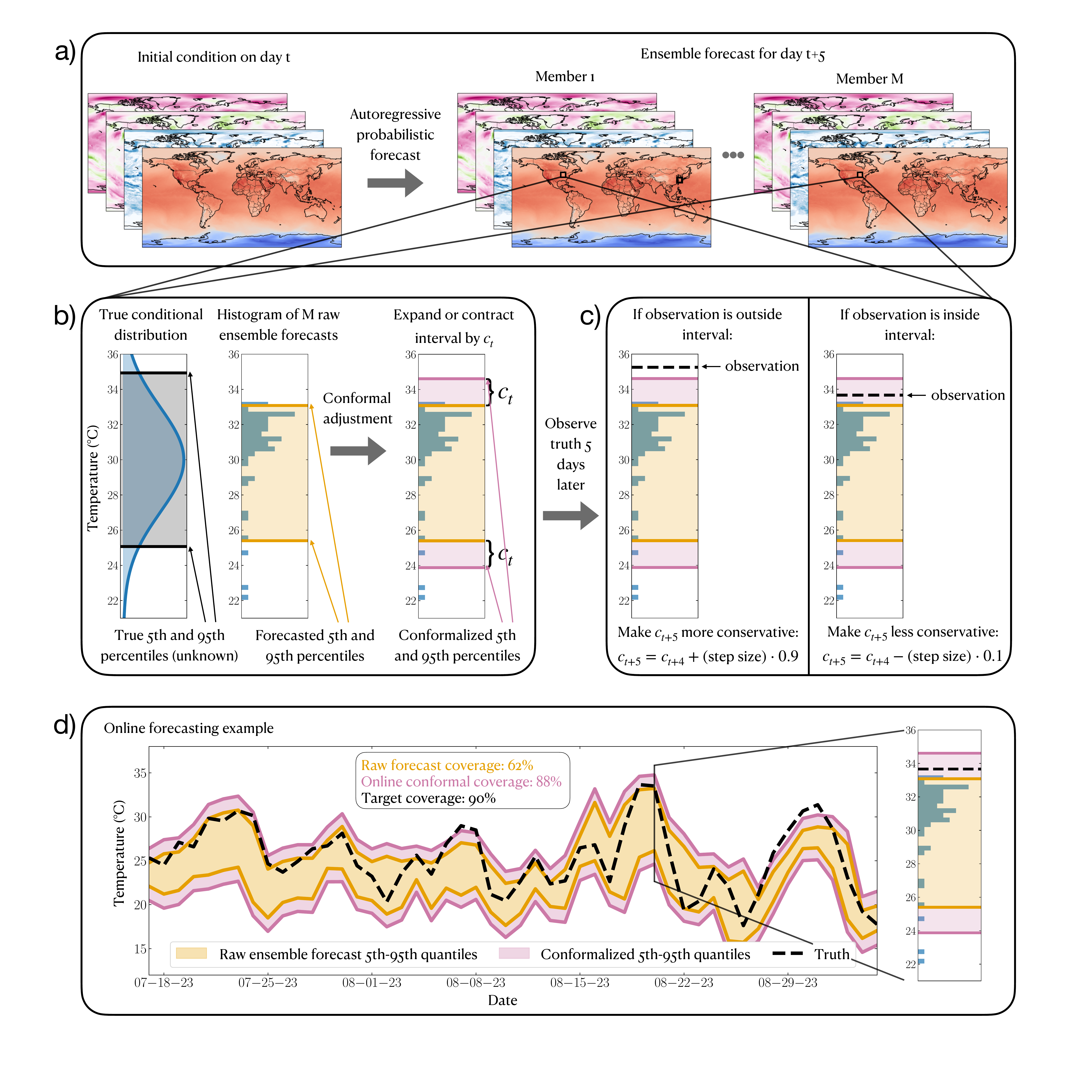}
\caption{
Schematic of the online adaptive conformal prediction framework for ensemble weather forecasts. We show $5$-day probabilistic prediction of $2$m temperature at a target coverage level of $90\%$ ($\alpha = 0.1$). a) We produce an ensemble $5$-day forecast of the global atmospheric state. b) From this global forecast, we extract the ensemble forecast at one location and for one variable (``Histogram of $M$ raw ensemble forecasts"). The ensemble members are samples from a distribution that approximates the unknown ``true conditional distribution" of the weather. Typically, there is a difference between the true and forecast $5$th percentiles (same for the $95$th percentiles); our goal is to correct the forecast percentiles over time using observations. We make a conformalized forecast by subtracting $c_t$ from the forecast $5$th percentile and adding $c_t$ to the forecast $95$th percentile. c) After $5$ days, we observe whether the truth lies inside or outside of the conformalized interval. We update $c_{t+5}$, which is the size of the conformal adjustment for the forecast to be issued at time $t+5$. d) Example time series showing the raw ensemble forecast quantiles, the conformalized forecast quantiles, and ground truth over several weeks of $2023$ near Chicago (including a heat wave). The size of the conformal adjustment changes over time. The adjoining histogram shows how the ensemble forecasts from panels (b) and (c) are integrated in the online framework.}
\label{fig:conformal_schematic}
\end{figure}

In this paper, we quantify the statistical coverage of several state-of-the-art probabilistic AI forecasting models on near-surface temperature and precipitation, showing that the outputs are uncalibrated, especially for extremes. We then apply a specific conformal prediction method, adaptive conformal prediction, as an online post-processing correction to forecast intervals, as visualized in Figure~\ref{fig:conformal_schematic} and described in Section \ref{sec:methods}. Our analysis in Section \ref{sec:results} shows that this method greatly improves forecast calibration with no lost skill in the CRPS and SSR. We discuss limitations and ways to further improve the method in Section \ref{sec:conclusions}.

\section{Methods}\label{sec:methods}

We use online conformal prediction to provide statistical coverage guarantees for probabilistic weather forecasts. At forecast initialization time $t$, let $X_t \in \mathbb{R}^d$ denote the atmospheric initial conditions, and $Y_{t+\tau} \in \mathbb{R}$ the true scalar quantity to be predicted at lead time $\tau$, such as $2$m temperature at a particular grid point. Let $\alpha \in (0,1)$ denote a miscoverage rate, so that $1-\alpha$ is the desired coverage rate. From an ensemble forecast, we extract lower and upper ensemble quantiles, denoted $\hat{q}_{\rm lo}(X_t)$ and $\hat{q}_{\rm hi}(X_t)$. For a target coverage level of $90\%$ ($\alpha=0.1$), these correspond to the empirical 5th and 95th percentiles of the ensemble. Our goal is to modify these raw quantiles so that, over a sequence of forecasts at times $t=1,\ldots,T$,
\begin{equation}
\frac{1}{T}\sum_{t=1}^T 
\mathbf{1}\{Y_{t+\tau} \in \hat{C}_t(X_t)\}
\approx 1-\alpha ,
\end{equation}
where $\hat C_t(X_t) = [\hat{q}_{\rm lo}(X_t), \hat{q}_{\rm hi}(X_t)]$ is a prediction interval using information available at time $t$, and $\mathbf{1}$ is the indicator function. In words, the fraction of forecasts whose prediction intervals contain the truth should approach the desired coverage level. 

\subsection{Online Conformal Prediction}\label{sec:ACI}

We explain an online conformal prediction procedure from \citeA{angelopoulos_conformal_2023} that tracks how much to expand or contract the raw ensemble intervals over time. Figure \ref{fig:conformal_schematic} visualizes the method. At each forecast time $t$, we form the conformalized prediction interval
\begin{equation}
\hat{C}_t(X_t)
=
\left[
\hat{q}_{\rm lo}(X_t) - c_t,\,
\hat{q}_{\rm hi}(X_t) + c_t
\right].
\end{equation}
Here $c_t$ is a time-varying padding term in the same physical units as the predicted variable, $Y_{t+\tau}$. If $c_t>0$, the interval is widened relative to the raw ensemble interval; if $c_t<0$, it is narrowed. Thus, rather than assuming that the raw ensemble quantiles yield perfect coverage, we learn a correction that adapts as forecast errors are observed.

After the lead time has elapsed, we observe whether the true value fell inside the interval. We define
\begin{equation}
\mathrm{err}_t
=
\mathbf{1}\{Y_{t+\tau} \notin \hat{C}_t(X_t)\},
\end{equation}
so that $\mathrm{err}_t=1$ when the interval misses the truth and $\mathrm{err}_t=0$ when it contains the truth. The padding is then updated according to
\begin{equation}
c_{t+\tau}
=
c_{t+\tau-1}
+
\eta(\mathrm{err}_t-\alpha),
\end{equation}
where the step size parameter $\eta>0$ controls how quickly the conformal correction responds to recent forecast performance. For the $5$-day forecasts shown schematically in Figure~\ref{fig:conformal_schematic}, this update is $c_{t+5} = c_{t+4} + \eta(\mathrm{err}_t-\alpha).$ This update differs from \citeA{angelopoulos_conformal_2023} in that it reflects the operational timing of the forecast: the outcome for the forecast issued on day $t$ is not known until day $t+5$, so the update can only affect forecasts issued after that verification time. The update uses $c_{t+4}$, the most recent padding value available, before incorporating the newly verified forecast. We prove convergence in Section S2.2.

This rule has an intuitive interpretation. If the interval misses the truth, then $\mathrm{err}_t=1$, and the padding increases by $\eta(1-\alpha)$, making future intervals more conservative. If the interval contains the truth, then $\mathrm{err}_t=0$, and the padding decreases by $\eta\alpha$, making future intervals less conservative. For a $90\%$ interval, $\alpha=0.1$: a miss increases $c_t$ by $0.9\eta$, while a successful coverage event decreases it by $0.1\eta$. Thus one miss is balanced by nine successful coverage events, matching the desired $10\%$ miscoverage rate.

The coverage guarantee follows directly from this adaptive update. Summing the update over $T$ verified forecasts gives
\begin{equation}\label{eq:guarantee}
\frac{1}{T}\sum_{t=1}^T \mathrm{err}_t
=
\alpha
+
\frac{c_{T+\tau}-c_{\tau}}{\eta T}.
\end{equation}
The second term describes how the empirical miscoverage rate differs from the intended rate $\alpha$. It approaches $0$ as $T\rightarrow\infty$, provided these $c_t$ terms are bounded, which is true if $Y$ is bounded. In such cases, the empirical miscoverage rate converges to $\alpha$. This is the adaptive conformal guarantee: if the original ensemble intervals are too narrow or too wide, the conformalized intervals adjust online toward the target coverage level \cite{gibbs_adaptive_2021}. In practice, the step size $\eta$ controls a tradeoff between stability and responsiveness. Smaller values produce gradual changes in intervals, while larger values allow quicker reaction to changing forecast skill and faster convergence, but possibly at the expense of stability. We discuss the role of $T$ in the Results, but note that empirical convergence to the target miscoverage rate occurs within days or weeks.

We employ the procedure separately for each lead time, variable, and grid point. In Section S2, we detail how we apply online conformal in our setting. We mirror the approach of \citeA{gibbs_adaptive_2021}, which does the adaptation in quantile space instead of variable space. This framework is more easily deployed across many locations in parallel, because the step size is dimensionless; updates become neutral to the climatology of the region.

\subsection{Ensemble Model Output Statistics}\label{sec:emos}

We also implement a post-processing baseline, the ensemble model output statistics (EMOS) method of \citeA{gneiting_calibrated_2005}. For temperature, we fit Gaussian distributions; for precipitation, we implement the left-censored generalized extreme value distribution and other modifications discussed in \citeA{scheuerer_probabilistic_2014} to account for non-Gaussianity and the possibility of a point mass at $0$. Implementation details are in Section S2.3.

\subsection{Models and Data}\label{sec:models}
Conformal prediction is agnostic to the model that produces the original quantiles $\hat{q}_{\rm lo}$ and $\hat{q}_{\rm hi}$: NWP, AI, or any other forecasting method could be used. Here, we evaluate our method on three state-of-the-art probabilistic weather forecasting models, two AI-based (GenCast and AIFS-ENS) and one hybrid (NeuralGCM), each of which produce ensemble forecasts \cite{price_probabilistic_2025, lang_aifs-crps_2026, kochkov_neural_2024}.  

GenCast is a conditional diffusion model that transforms Gaussian noise into a forecast. Depending on the year, we have $52$ or $56$ ensemble members. AIFS-CRPS is a transformer-based model, trained with a CRPS-based loss. We use AIFS-ENS, a version fine-tuned on operational Integrated Forecasting System data, and generate 25 ensemble members.  NeuralGCM, which has a dynamical core and machine-learned closures, has a stochastic version \cite{yuval_neural_2026} fine-tuned to match satellite-based precipitation observations from the Integrated Multi-satellitE Retrievals for GPM (IMERG) dataset \cite{huffman_gpm_2023}. We generate $51$ ensemble members. For each model, we estimate quantiles by linearly interpolating between ensemble members. As ground truth, we take the ERA5 reanalysis dataset on which these models were trained \cite{hersbach_era5_2020}. The only exception is NeuralGCM precipitation, for which we use IMERG. See Section S3 for further details.

In the Results, we also evaluate the performance on extremes. To define extremes, we select as a threshold the 95th percentile of climatology, calculated for each spatial location and calendar date from ERA5 reanalysis data spanning 1979--2018 (except for NeuralGCM precipitation, for which we use IMERG data from 2000--2018). The coverage on extremes is a type of conditional coverage, and the theoretical guarantee does \textit{not} hold in this setting \cite{foygel_barber_limits_2021}. 

\section{Results}\label{sec:results}
 
For each AI model, we evaluate globally on $12$-hour total precipitation and near-surface temperature (two-meter temperature for GenCast and AIFS-ENS, and temperature at $1000$ hPa for NeuralGCM). We calibrate each model at many different $\alpha$ levels to improve the entire forecast distribution.

We first consider a lead time of $\tau = 5$ days and $\alpha = 0.1$ ($90\%$ coverage). For each variable and grid point (indexed by $i$), we report the fraction of verification times without a miss as the empirical coverage:
\begin{equation}\label{eq:cov}
{\rm coverage}_i = \frac{1}{T}\sum_{t=1}^T (1-{\rm err}_{i,t}),
\end{equation}
where $t$ indexes the forecast dates available over the test period from $2022$ to $2024$. Here, a perfectly calibrated forecast would have empirical coverage of $90\%$.

\begin{figure}
\centering
\noindent\includegraphics[width=0.8\textwidth]{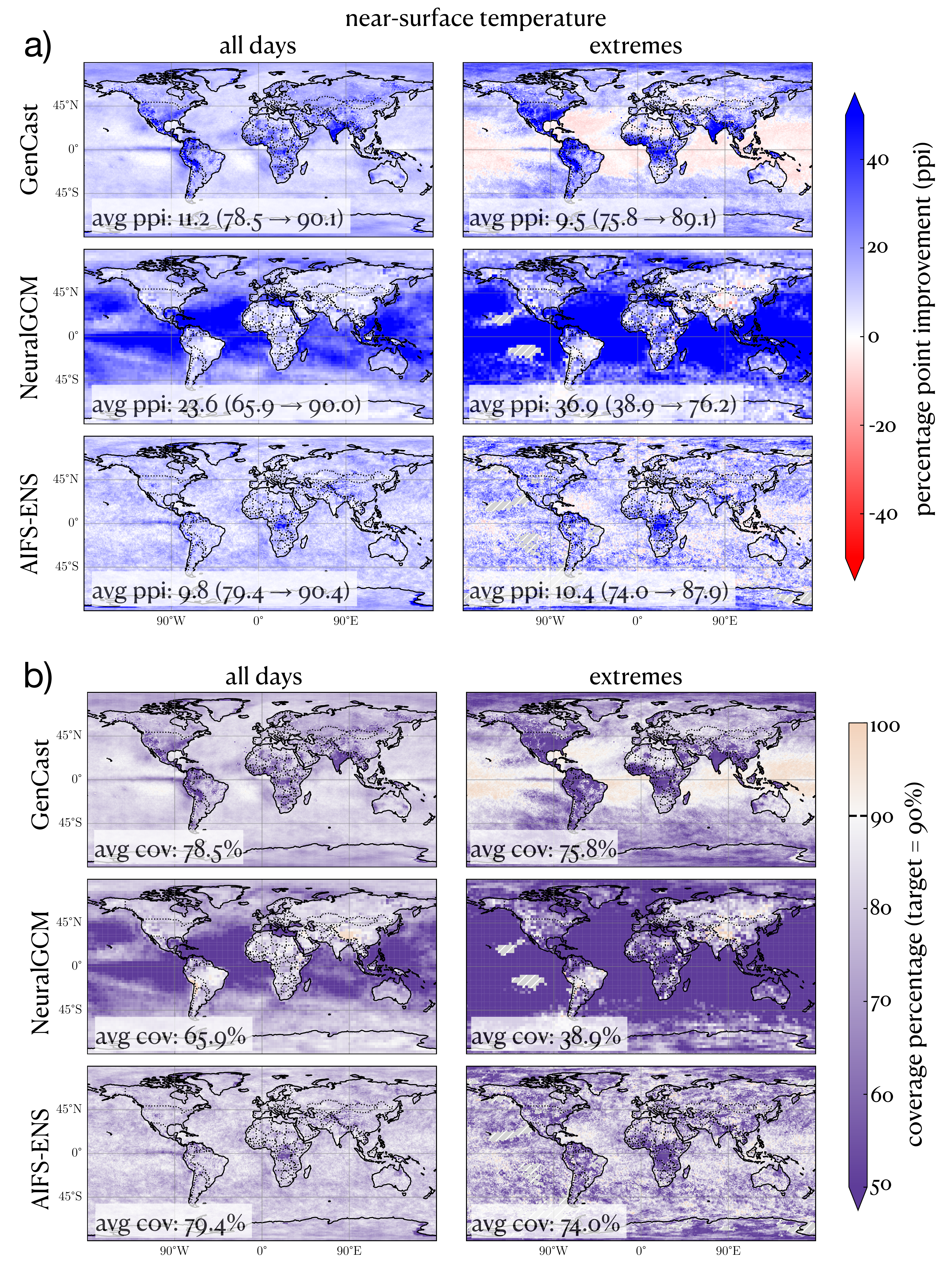}
\caption{a) Left: Spatial map of coverage improvement on near-surface temperature at a target level of $90\%$, averaged over the test period from $2022$--$2024$. Values are reported in percentage point improvement (ppi), as defined in Equation \eqref{eq:ppi}. We track independent $c_t$ values for each grid point. Blue indicates regions where the conformal adjustment improves coverage. The number in the lower left-hand corner is the area-weighted global average ppi. In parentheses is how the area-weighted global average empirical coverage changed between the original and conformalized models. (Their difference may not equal the average ppi because the global average does not commute with the absolute values in Equation \eqref{eq:ppi}.) Right: The same, but conditioned on the ground truth being extreme---above the 95th percentile of climatology for the given location and calendar date. Gray hatching indicates locations where 10 or fewer days exceeded the threshold over the test period. Hatching differences across models are due to differences in spatial or temporal resolution. b) Left: Spatial map of the empirical coverage of the original ensemble forecasts. The number in the lower left-hand corner is the area-weighted global average coverage. Right: The same, but conditioned on the ground truth being extreme.}
\label{fig:temp}
\end{figure}

\begin{figure}
\centering
\noindent\includegraphics[width=0.8\textwidth]{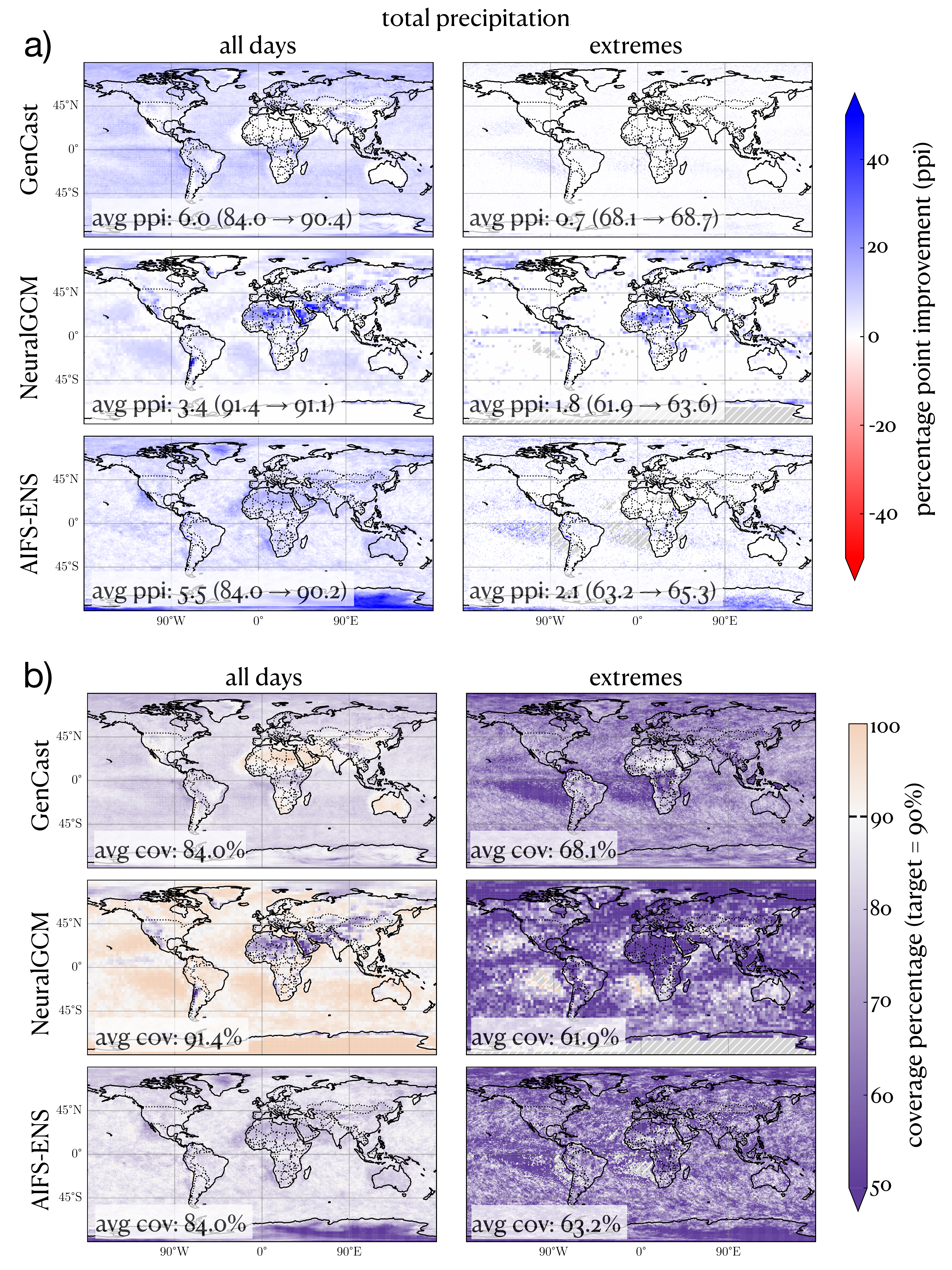}
\caption{The same as Figure~\ref{fig:temp}, but for total precipitation.}
\label{fig:precip}
\end{figure}

We calculate the empirical coverage for the original ensemble and the conformalized forecast intervals, and then quantify the coverage gain resulting from conformal prediction with the following value, which we call percentage point improvement (ppi):
\begin{equation}
{\rm ppi}_{i} := \left|{\rm raw\ ensemble\ coverage}_i - 0.9\right| - \left|{\rm conformalized\ coverage}_i - 0.9\right|.
\label{eq:ppi}\end{equation}
Starting with near-surface temperature, the left of Figure~\ref{fig:temp}a shows the global spatial distribution of these values over the test period. The right is the same, but only for days for which the truth exceeded the 95th percentile of climatology, as defined above.
Figure~\ref{fig:precip}a is the same, but for total precipitation.

Conformal prediction improves coverage in almost all cases, as the vast majority of grid points are blue. To understand the spatial variability, we turn to Figures~\ref{fig:temp}b and \ref{fig:precip}b, which plot the empirical coverage values (Equation \eqref{eq:cov}) of the raw forecasting models. Grid points for which there is little gain due to conformalization correspond well with those for which the original forecast already achieved good coverage. Then Figures S1 and S2 show that, on average, the conformalized forecasts achieve the desired marginal coverage almost exactly, as expected due to the convergence guarantee in Section \ref{sec:ACI}. In practice, convergence to within $0.01$ of the target coverage rate occurs within days, and to within $0.001$ after about a month.
There are grid points for which there remains over-coverage in Figure~S2, e.g., NeuralGCM over Antarctica. The original $5$th and $95$th percentiles and ground truth are often all $0$, so we cannot decrease coverage without sometimes outputting empty prediction intervals (i.e., $\hat{C}_t(X_t) = \emptyset$), which we decide not to do. 

Next we discuss the spatial variability of coverage. As examples seen in Figures~\ref{fig:temp} and \ref{fig:precip}, forecasts are improved greatly for NeuralGCM total precipitation over the Sahara, GenCast 2m temperature over the Andes and India, and AIFS-ENS 2m temperature over central Africa. Also, the original coverage on near-surface temperature is better over the ocean than the land for GenCast, the opposite is true for NeuralGCM, and the coverage is similar over land and ocean for AIFS-ENS. It is worthwhile to further explore the regional deficiencies of different models, as globally averaged metrics fail to capture spatial heterogeneity in model skill or calibration. Any relationship between forecast physics and coverage remains to be explored.

Turning to the results for extremes (above the 95th percentile; Section \ref{sec:models}), we note that the raw forecast coverage is uniformly worse on extremes than overall (compare the left and right columns in Figures \ref{fig:temp}b and \ref{fig:precip}b). The spatial distribution of forecast coverage also differs. The reduced quality of the raw ensemble quantiles on extremes can allow conformalization to lend greater improvement on extremes than typical weather, especially noticeable in, for example, NeuralGCM's near-surface temperature. While in all examples the globally averaged coverage on extremes improves after conformalization, there is room for further improvement, as seen on the right-hand side of Figures S1 and S2, especially for precipitation extremes. 

\begin{figure}
\noindent\includegraphics[width=\textwidth]{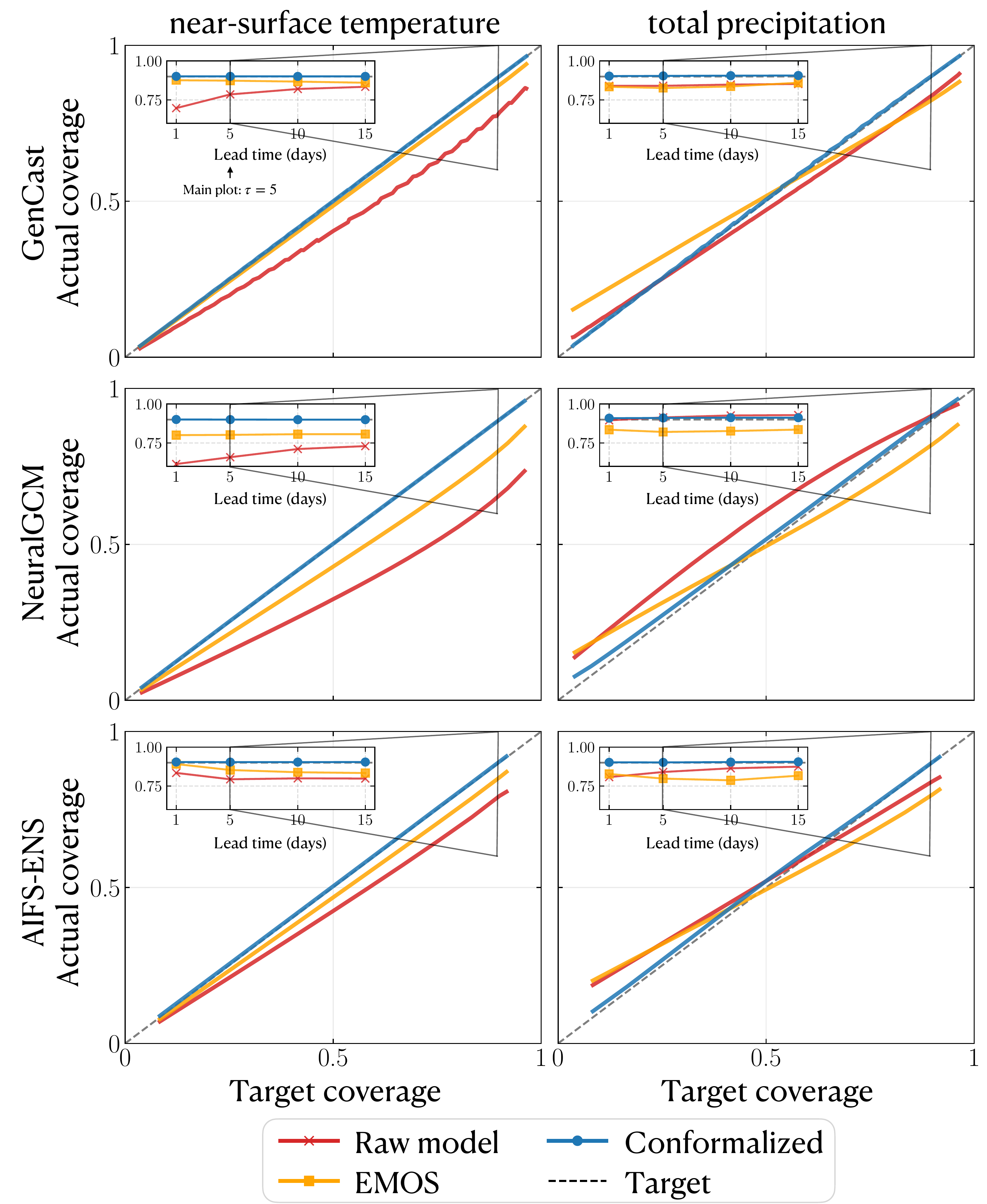}
\caption{Reliability diagram showing coverage across different target levels at a lead time of $\tau=5$ days, including a comparison against the EMOS baseline (see Section \ref{sec:emos}). Each point plots a specified target coverage level against the area-weighted globally averaged coverage over the test period, $2022$--$2024$. The black dashed line indicates a perfect model, red the original model, gold the EMOS-adjusted intervals, and blue the conformalized intervals. The conformalized predictions have reliability lines closely aligned with that of the perfect model. The inset panels show coverage across various lead times $\tau$ at the $90\%$ target coverage level.}
\label{fig:all_lev}
\end{figure}

The results thus far focus on $\alpha=0.1$ and $\tau=5$ days, but, as shown in the reliability diagram in Figure~\ref{fig:all_lev}, we also achieve the desired coverage across different target levels and lead times. In most cases the original ensembles undercover, but sometimes they overcover (e.g., NeuralGCM's total precipitation at target coverage below $90\%$); conformal prediction corrects both. In Figure~\ref{fig:all_lev}, we also compare against a baseline, EMOS, which generally improves calibration. But as conformal prediction comes with a statistical coverage guarantee, it produces near-perfect reliability diagrams. Additionally, the insets show that the method works across lead times from $1$ to $15$ days.

Better coverage might be achieved at the cost of unreasonable increases to the interval width. However, we demonstrate that this is not the case: the CRPS and SSR are improved or barely affected. See Section S1 for definitions of these metrics and how we adapt them to quantile-based forecasts \cite{gneiting_comparing_2011}. Figure S3 shows that the SSR generally improves after conformalization, and Figure S4 shows that the CRPS remains almost unchanged. So, when conformal prediction increases the widths of prediction intervals, it is in a principled way---if the interval was not wide enough before. 

\section{Summary and Conclusions}\label{sec:conclusions}

In this work, we demonstrate that online conformal prediction is an effective calibration step for global weather forecasting. The raw ensemble forecasts of several AI weather models lack calibration---generally producing overly narrow ranges of forecast values, which manifests in lower statistical coverage than desired. We correct the coverage deficiency at all target coverage levels with online conformal prediction, endowing outputs with a statistical guarantee without sacrificing skill on CRPS or SSR. The online framework adds value over traditional conformal prediction because the adjustments, $c_t$, can react to distributional changes, like seasonal differences or long-term trends \cite{gibbs_adaptive_2021}.

We also evaluate forecast coverage on rare extreme events. AI models have been shown to struggle to predict extreme events, especially gray swans, for which UQ becomes particularly important \cite{sun_can_2025, sun_predicting_2025, zhang_physics-based_2026}. Indeed, the models we evaluate do not correctly represent the likelihood of extremes. The conformalized forecasts improve, but do not perfect, coverage of extreme temperature; for extreme precipitation, improvement is marginal. Tracking separate upper and lower conformal corrections \cite{romano_conformalized_2019} would likely help, but is left for future work. More broadly, developing conformal prediction methods tailored for extremes remains an important challenge. A natural first step is to build upon the uncertainty-aware framework developed by \citeA{rossellini2024integrating}. 

We use conformal prediction as a post-processing method, and, as a baseline, we take a classic method, EMOS. Newer techniques exist, but operational state-of-the-art post-processing methods in NWP have become highly specialized \cite{hewson_low-cost_2021, roberts_improver_2023}. Methods like these, or EMOS executed after a more robust parameter search, might provide a stronger baseline (though they will still lack statistical guarantees). However, conformal prediction works out-of-the-box and requires little to no parameter tuning. Our experimental verification that it achieves target coverage demonstrates its utility. 

Next, we discuss some relevant aspects of the algorithm. Equation~\eqref{eq:guarantee} guarantees convergence to the desired miscoverage rate $\alpha$ as $T\rightarrow\infty$. Empirically, online conformal improves coverage even for modest values of $T$, corresponding to less than a month. Another important factor is the number of ensemble members present in the original forecast, $M$. As $M$ increases, the raw forecast coverage may improve, but even as $M\rightarrow\infty$, the empirical forecast distribution still differs from the unknown, ground truth distribution due to structural errors, so we cannot produce well-calibrated forecasts by simply sampling massive ensembles from AI models. Even if the original model is poor, some amount of interval inflation can always recover good coverage, but the intervals will be large. Conformal prediction learns a data-driven correction directly from observed forecast errors, achieving good coverage without naive inflation. 

Our implementation of conformal prediction comes with natural limitations. It only corrects the variance; correcting the bias in tandem could further improve performance. Another drawback is that we must fit a different conformal adjustment for each location, lead time, variable, and quantile. On the positive side, this means the computational cost of the real-time update is easily parallelizable and minimal for each---simply computing a quantile and adding and removing a scalar from a small list. However, this breaks spatial covariance and cross-variable interactions. A good avenue of future work could be reintroducing dependence structure; methods in the literature address this \cite{clark_schaake_2004, schefzik_uncertainty_2013}, but the distribution-free guarantee would be lost.

%
%

\section*{Open Research Section}
We obtain GenCast forecasts from Google DeepMind's WeatherNext Gen, which produces forecasts with an operational version of GenCast (see \url{https://developers.google.com/earth-engine/datasets/catalog/projects_gcp-public-data-weathernext_assets_126478713_1_0}). We generated the forecasts for Google Research's NeuralGCM (2.8° stochastic precipitation version) and ECMWF’s AIFS-ENS; the model weights are freely available from \url{https://neuralgcm.readthedocs.io/en/latest/checkpoints.html} and \url{https://huggingface.co/ecmwf/aifs-ens-1.0}.
ERA5 reanalysis is freely available from the Copernicus Climate Data Store \cite{hersbach2023era5, hersbach2023era5b}.
IMERG data can be accessed from \url{https://doi.org/10.5067/GPM/IMERG/3B-HH/07} \cite{huffman_gpm_2023}. Code that produces the results in this paper will be provided upon acceptance.

\section*{Conflict of Interest declaration}
The authors declare there are no conflicts of interest for this manuscript.

\section*{Acknowledgments}
We thank Adam Marchakitus and Bing Gong for helping produce the original forecasts. A.A.\ is supported by NSF GRFP-2140001. P.H.\ and R.W.\ are grateful for support from NSF AGS-2531264 and AFSOR FA9550-24-1-0327, respectively. This work also received support from the University of Chicago's Data Science Institute, the Institute of Climate and Sustainable Growth, and the Laude Foundation. Computational resources were provided by NSF ACCESS (ATM170020), NCAR's CISL (UCHI0014), and the University of Chicago Research Computing Center.


\bibliography{references}

%
%

\clearpage
\appendix

\setcounter{figure}{0}
\setcounter{table}{0}
\setcounter{equation}{0}
\setcounter{section}{0}

\renewcommand{\thefigure}{S\arabic{figure}}
\renewcommand{\thetable}{S\arabic{table}}
\renewcommand{\theequation}{S\arabic{equation}}
\renewcommand{\thesection}{S\arabic{section}}

\begin{center}
    \LARGE \textbf{Supporting Information} \\[0.5em]
\end{center}

\section*{Contents of this file}
\begin{enumerate}
\item Texts S1 to S3
\item Figures S1 to S4
\end{enumerate}

\section*{Introduction}

Text S1 contains precise definitions of the probabilistic metrics used for evaluation. Text S2 provides algorithmic implementation details for online conformal prediction and the ensemble model output statistics method, as well as a proof that online conformal prediction converges with delayed updating. Text S3 explains technical details of the data used.

\section{Probabilistic Metrics}

First, we spell out notation, following \citeA{price_probabilistic_2025} closely. All notation is for a particular variable and lead time. Let $x_{i,t}^m$ be the value of the $m$th of $M$ ensemble members from initialization time $t = 1,\ldots, T$ at grid point $i \in G$. Let $y_{i,t}$ be the verifying observation. Let $a_i$ denote the area of the $i$th grid cell. Two useful derived quantities in the following exposition are the ensemble mean, $\overline{x}_{i,t}$, and the unbiased sample variance of the ensemble members,
\[s_{i,t}^2 = \frac{1}{M-1} \sum_{m=1}^M \left(\overline{x}_{i,t} - x_{i,t}^m\right)^2.\]

Standard implementations of probabilistic metrics in weather forecasting rely on an ensemble to characterize the predictive distribution. However, conformal prediction inherently outputs prediction intervals instead of ensemble members. In the following subsections, we also explain how we calculate metrics in this case, which requires running the algorithm at many target miscoverage levels $\alpha$.

\subsection{Continuous-Ranked Probability Score}
Fix location $i$ and initialization time $t$ and let $F(x)$ be the empirical cumulative distribution function (CDF) of the ensemble $x_{i,t}^m$, $m=1,\ldots,M$. The continuous ranked probability score (CRPS), in the negative orientation \cite{gneiting2007strictly}, is defined as 
\[\mathrm{CRPS}(F,y_{i,t}) = \int_{-\infty}^\infty \left(F(z) - \mathbf{1}\{z \ge y_{i,t}\}\right)^2\, dz.\]
This is the squared $L^2$ distance between the empirical and actual CDFs, where the actual CDF is a Heaviside function at the value of the verifying observation, $y_{i,t}$. We then average this value over all initialization times and grid points under consideration, obtaining the CRPS:
\[\mathrm{CRPS} = \frac{1}{T} \sum_{t=1}^T \frac{1}{|G|} \sum_{i} a_i \mathrm{CRPS}(F,y_{i,t}).\]

When working with quantile forecasts instead of ensemble members, we approximate the integral with a different formula. Following the presentation in \citeA{gneiting_comparing_2011}, let $F$ be the CDF of the forecast and $F^{-1}(\alpha)$ the quantile forecast at level $\alpha$. Then, with the quantile score $\mathrm{QS}_{\alpha}(F^{-1}(\alpha),y) = 2(\mathbf{1}\{y\le F^{-1}(\alpha)\} - \alpha)(F^{-1}(\alpha) - y)$,
\[\mathrm{CRPS}(F,y_{i,t}) = \int_{0}^1 \mathrm{QS}_{\alpha}(F^{-1}(\alpha),y_{i,t})\, d\alpha.\]
We discretize this integral with the available $\alpha$ values.

\subsection{Spread-Skill Ratio}

The spread-skill ratio compares, roughly, the ensemble spread to average ensemble mean error. The first term to consider is the average ensemble variance,
\[\mathrm{AvgEnsVar} = \frac{1}{T}\sum_{t=1}^T \frac{1}{\left|G\right|} \sum_{i} a_is_{i,t}^2,\]
and the mean squared error (MSE) of the ensemble mean is
\[\mathrm{EnsMeanMSE} = \frac{1}{T} \sum_{t=1}^T \frac{1}{\left|G\right|} \sum_{i} a_i\left(\overline{x}_{i,t} - y_{i,t}\right)^2\]
Some authors bias-correct this term to obtain a fair estimate \cite{price_probabilistic_2025}, and we also do this. This bias-corrected value is 
\[\mathrm{EnsMeanMSE}_{\textrm{fair}} = \mathrm{EnsMeanMSE} - \frac{\mathrm{AvgEnsVar}}{M}.\]
Then the SSR is
\[{\mathrm{SSR}} = \sqrt{\frac{\mathrm{AvgEnsVar}}{\mathrm{EnsMeanMSE}_{\textrm{fair}}}}.\]

If the verifying observation is second-order exchangeable with the ensemble members, the SSR equals $1$ in expectation \cite{fortin_why_2014}. Assuming little bias in the forecast, a forecast that is underdispersive (overdispersive) on average will have $\mathrm{SSR} <$ ($>$) $1$.

When working with quantiles instead of ensemble members, we cannot directly estimate the ensemble mean and variance. Instead, we estimate the mean as $\mathbb{E}[X] = \int_0^1 F^{-1}(\alpha)\, d\alpha$ and the variance as $\mathrm{Var}[X] = \int_0^1(F^{-1}(\alpha) - \mathbb{E}[X])^2\, d\alpha$ (assuming finite variance).

\section{Algorithmic Details}
\setcounter{section}{2}
\setcounter{subsection}{0}

\subsection{Online Conformal Prediction}
For each model, we use data from $2021$--$2024$. Forecasts initialized in $2021$ serve as pure calibration data. Then we step through all forecast initialization dates in $2022$--$2024$, using the most recent year's worth of nonconformity scores as the calibration data with which to calculate the conformal adjustment. We set the step size $\eta = 0.01$, following \citeA{gibbs_adaptive_2021}. 

\subsection{Online Conformal Prediction Delayed Updating Proof}

\noindent\textbf{Proposition S1.}
Fix $\alpha \in [0,1]$, $\eta>0$, and $\tau \in \mathbb N_{>0}$. Suppose $Y_t \in [-b/2,b/2]\;\; \forall t\geq 1$. Suppose we use the $\tau$-step delayed update rule 
$$
    c_{t+\tau} = c_{t+\tau-1} + \eta (\mathrm{err}_t - \alpha),
$$
where $\mathrm{err}_t = \mathbf{1}\{Y_{t+\tau} \notin [\hat{q}_{\rm lo}(X_t) - c_t, \hat{q}_{\rm hi}(X_t)+c_t]\}$. Assume $\hat{q}_{\rm lo}(\cdot), \hat{q}_{\rm hi}(\cdot) \in [-b/2, b/2]$. In the case that $\hat{q}_{\rm lo}(X_t) - c_t > \hat{q}_{\rm hi}(X_t)+c_t$, we adopt the convention that $\mathrm{err}_t = 1$.

If we initialize $c_1,\dots,c_\tau =0$, then
$$
 \left|\frac{1}{T}\sum_{t=1}^T (\mathrm{err}_t - \alpha)\right| \leq \frac{b+\tau\eta}{\eta T}.
$$
This recovers the guarantee of \citeA[Proposition 1]{angelopoulos_conformal_2023}, who study the case of $\tau=1$.

\par\noindent\textit{Proof.}

Observe that
$$
    \frac{1}{T}\sum_{t=1}^T c_{t+\tau} = \frac{1}{T}\sum_{t=1}^T c_{t+\tau-1} + \frac{1}{T}\sum_{t=1}^T \eta (\mathrm{err}_t - \alpha).
$$

By cancellation, we get
$$
    \frac{1}{\eta T}(c_{T+\tau} - c_{\tau}) = \frac{1}{T}\sum_{t=1}^T (\mathrm{err}_t - \alpha).
$$

We know that $\mathrm{err}_t = 0$ if $c_t \geq b$, and, similarly, $\mathrm{err}_t=1$ if $c_t \leq -b$, by the boundedness assumption on $Y$.

We assert that $-b-\tau\eta\alpha \leq c_t \leq b + \tau\eta(1-\alpha)$.

The upper bound holds trivially if $c_t \leq b$. So we suppose $c_t > b$. Let $u\leq t$ refer to the last time index for which $c_u \leq b$. We know such an index must exist since $c_1,\dots, c_\tau = 0$ by assumption. For all indices $s$ such that $u+\tau < s \leq t$, we must have $c_{s-\tau} > b$, by the definition of $u$. Therefore, $\mathrm{err}_{s-\tau} = 0$ and \[
c_s - c_{s-1} = \eta(\mathrm{err}_{s-\tau}- \alpha) \leq 0.
\]

Thus, after time $u$, only the first $\tau$ updates yield increases. Each such increase is at most $\eta(1-\alpha)$. Therefore
\[
c_t
\leq
c_u+\tau\eta(1-\alpha)
\leq
b+\tau\eta(1-\alpha).
\]
This proves the upper bound. The lower bound proof follows from an analogous argument.

Using these bounds and the fact that $c_\tau=0$ by assumption, we have
$$
    \left |\frac{1}{T}\sum_{t=1}^T (\mathrm{err}_t - \alpha)\right| = \left|\frac{1}{\eta T}(c_{T+\tau} - c_{\tau})\right| \leq  \frac{b+\tau\eta}{\eta T}.
$$

\subsection{Ensemble Model Output Statistics}\label{sec:emos_si}

As in the original \citeA{gneiting_calibrated_2005} paper, we fit the following Gaussian predictive distribution at each grid point:
\[\mathcal{N}\left(a+b\overline{X}, c+ dS^2\right),\]
where $a$, $b$, $c$, and $d$ are learned coefficients, $\overline{X}$ is the forecast mean, and $S^2$ is the forecast variance. We use a single coefficient $b$ for the ensemble mean $\overline{X}$, rather than fitting separate coefficients for each ensemble member, because we work with a single forecasting system with exchangeable ensemble members. 

We also implement a precipitation-specific version, the left-censored generalized extreme value (GEV) distribution proposed by \citeA{scheuerer_probabilistic_2014}. This model characterizes precipitation with a CDF of the form
\[\widetilde{G}(y) = \begin{cases}
    G(y), &y \ge 0 \\
    0, & y<0,
\end{cases}\]
where $G(y)$ is the CDF of the GEV distribution,
\[G(y) = \begin{cases}
    \exp{\left[-\left\{1 + \xi\left(\frac{y-\mu}{\sigma}\right)\right\}^{-1/\xi}\right]}, &\xi \ne 0 \\
    \exp{\left[-\exp\left\{\left(-\frac{y-\mu}{\sigma}\right)\right\}\right]}, &\xi = 0,
\end{cases}\]
with $G(y) := 1$ for $\xi <0$ and $y > \mu - \sigma/\xi$, and $G(y) := 0$ for $\xi >0$ and $y < \mu - \sigma/\xi$. \citeA{scheuerer_probabilistic_2014} restricts $\xi \in (-0.278,1)$, and over this range, the mean of the GEV distribution is
\[m = 
\begin{cases}
\mu + \sigma \frac{\Gamma(1-\xi) - 1}{\xi}, &\xi \ne 0 \\
\mu + \sigma\gamma, &\xi = 0
\end{cases},\]
where $\Gamma$ is the gamma function and $\gamma$ is the Euler–Mascheroni constant.

At each grid point $i$, let $\overline{X}$ be the forecast mean as before, $\overline{\mathbf{1}_{X = 0}}$ be the fraction of ensemble members predicting zero precipitation, and $\mathrm{MD}(X) = \frac{1}{M^2}\sum_{m,m'}\left|x_{i,t}^m - x_{i,t}^{m'}\right|$ be the ensemble mean difference. Then, we connect parameters $m_i$ and $\sigma_i$ to the ensemble output via
\begin{align}
m_i &= \alpha_0 + \alpha_1\cdot \overline{X} + \alpha_2 \cdot \overline{\mathbf{1}_{X=0}} \\
\sigma_i &= \beta_0 + \beta_1 \cdot \mathrm{MD}(X).
\end{align}
The coefficients $\alpha_0$, $\alpha_1$, $\alpha_2$, $\beta_0$, $\beta_1$, and $\xi$ must be estimated at each spatial location. To make the parameter estimation more robust, we implement spatial pooling, at each spatial location expanding the training data to include observations from the surrounding $3\times3$ block of grid points.

In both cases, we estimate the parameters by minimizing the CRPS over the most recent $30$ days of verified observations.

\section{Data}

GenCast forecasts are initialized daily at 0Z, and we subsample from the native $0.25^\circ$ grid to a $1^\circ$ grid. There are $56$ ensemble members available for the years $2021$, $2022$, and $2023$, and $52$ ensemble members for $2024$. NeuralGCM forecasts have $51$ ensemble members at $2.8^\circ$ resolution. We run the model from initial conditions at 0Z every $3$ days. We generate AIFS-ENS forecasts with $25$ ensemble members twice weekly, initialized at 0Z, and subsample from the native $0.25^\circ$ grid to a $1^\circ$ grid. 

For all models, we use data from $2021$--$2024$. This time period is completely out of sample for GenCast and NeuralGCM. AIFS-ENS, however, was fine-tuned on operational Integrated Forecasting System analysis data that includes $2021$--$2023$, so these results are not strictly out of sample.

When we evaluate on ``near-surface temperature", this is $2$m temperature for GenCast and AIFS-ENS, and temperature at $1000$ hPa for NeuralGCM (since this model does not have $2$m temperature). For ``precipitation", we evaluate on total $12$-hour accumulated precipitation, for all models.

%
%
\clearpage

\begin{figure}
\noindent\includegraphics[width=\textwidth]{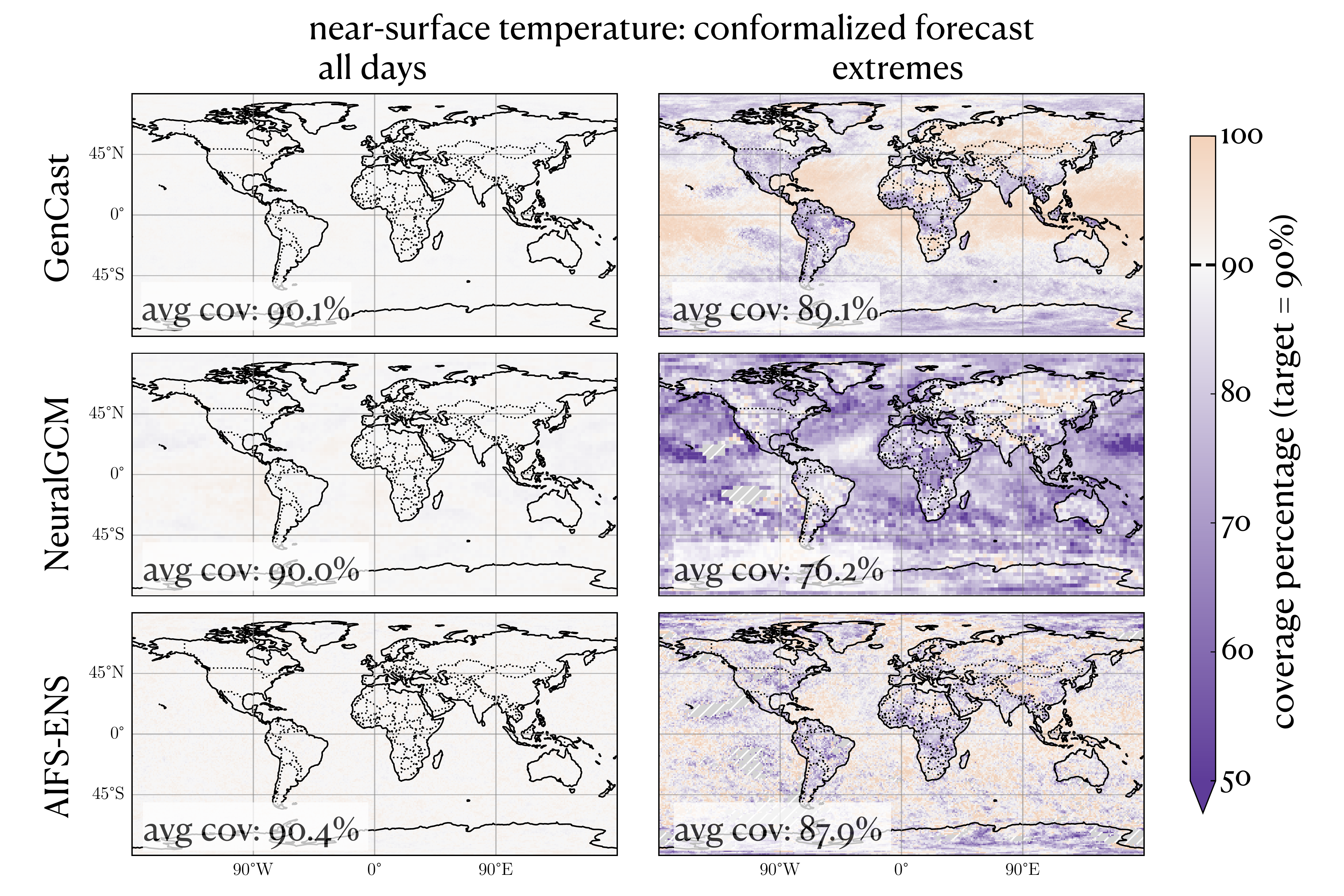}
\caption{The same as Figure 2b, but for the conformalized forecasts of near-surface temperature.}
\label{fig:conf_temp}
\end{figure}

\begin{figure}
\noindent\includegraphics[width=\textwidth]{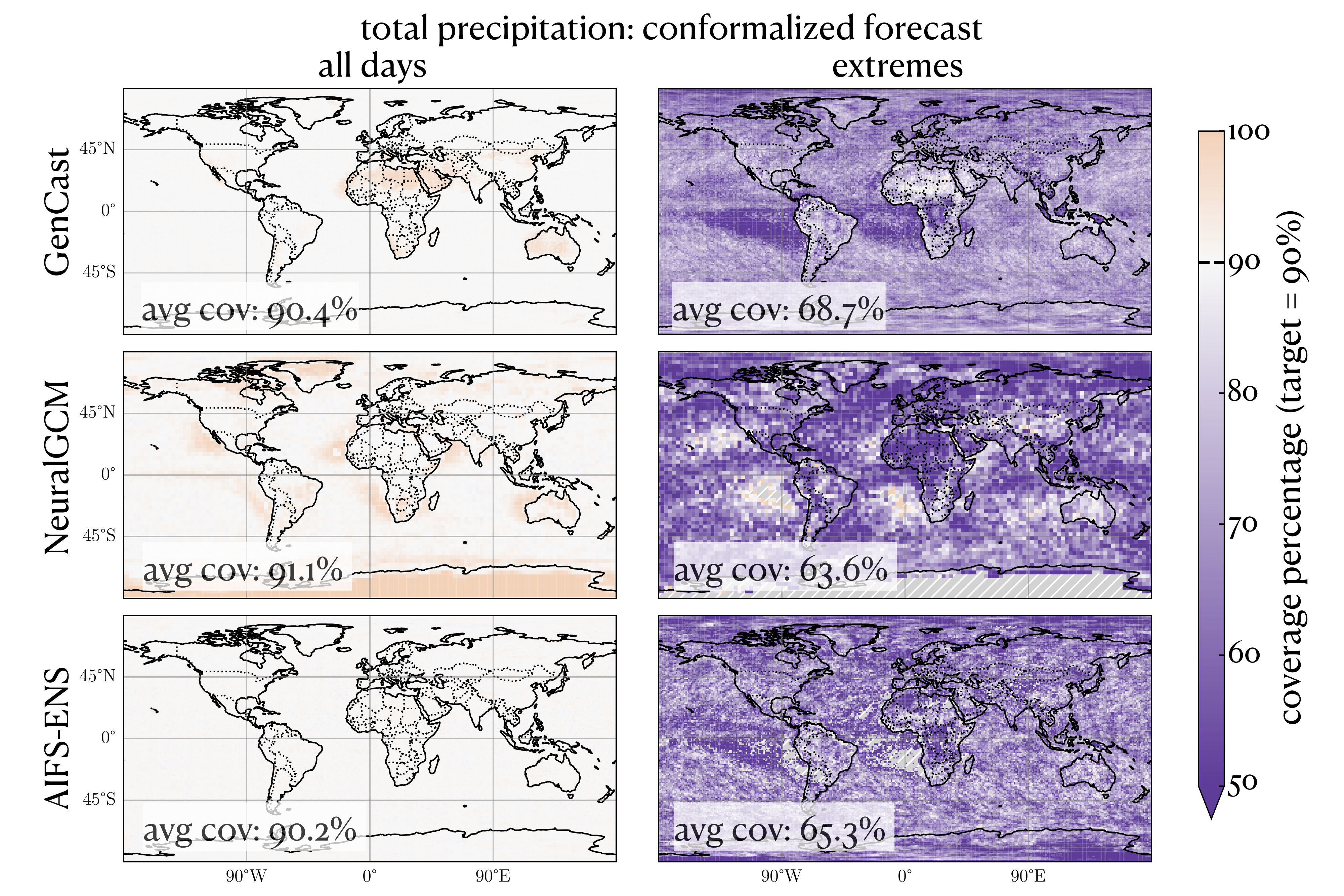}
\caption{The same as Figure 3b, but for the conformalized forecasts of total precipitation.}
\label{fig:conf_precip}
\end{figure}

\begin{figure}
\noindent\includegraphics[width=\textwidth]{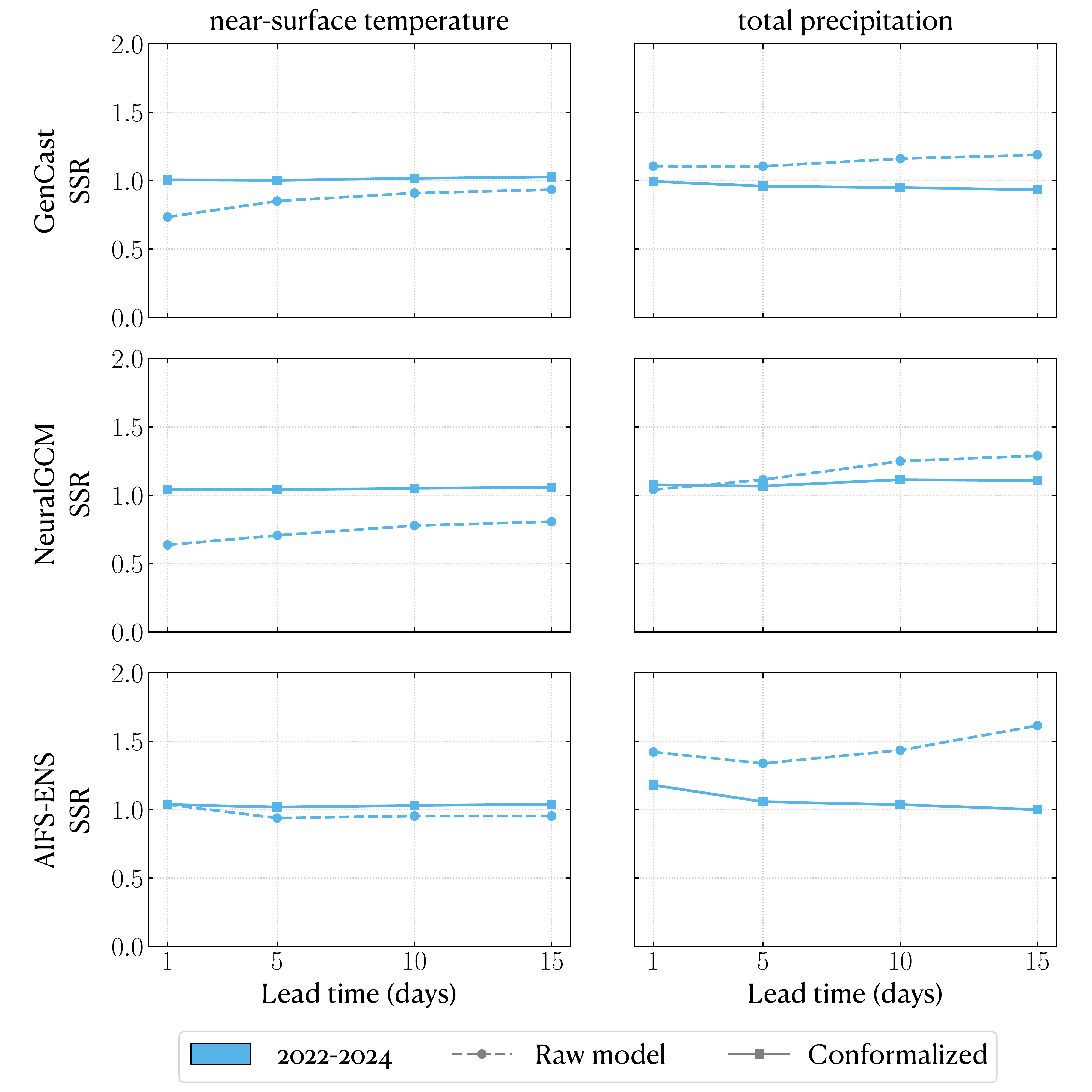}
\caption{The SSR as a function of lead time. The conformalized model (solid lines) generally performs better (i.e., has values closer to 1) than the raw ensembles (dashed lines).}
\label{fig:ssr}
\end{figure}

\begin{figure}
\noindent\includegraphics[width=\textwidth]{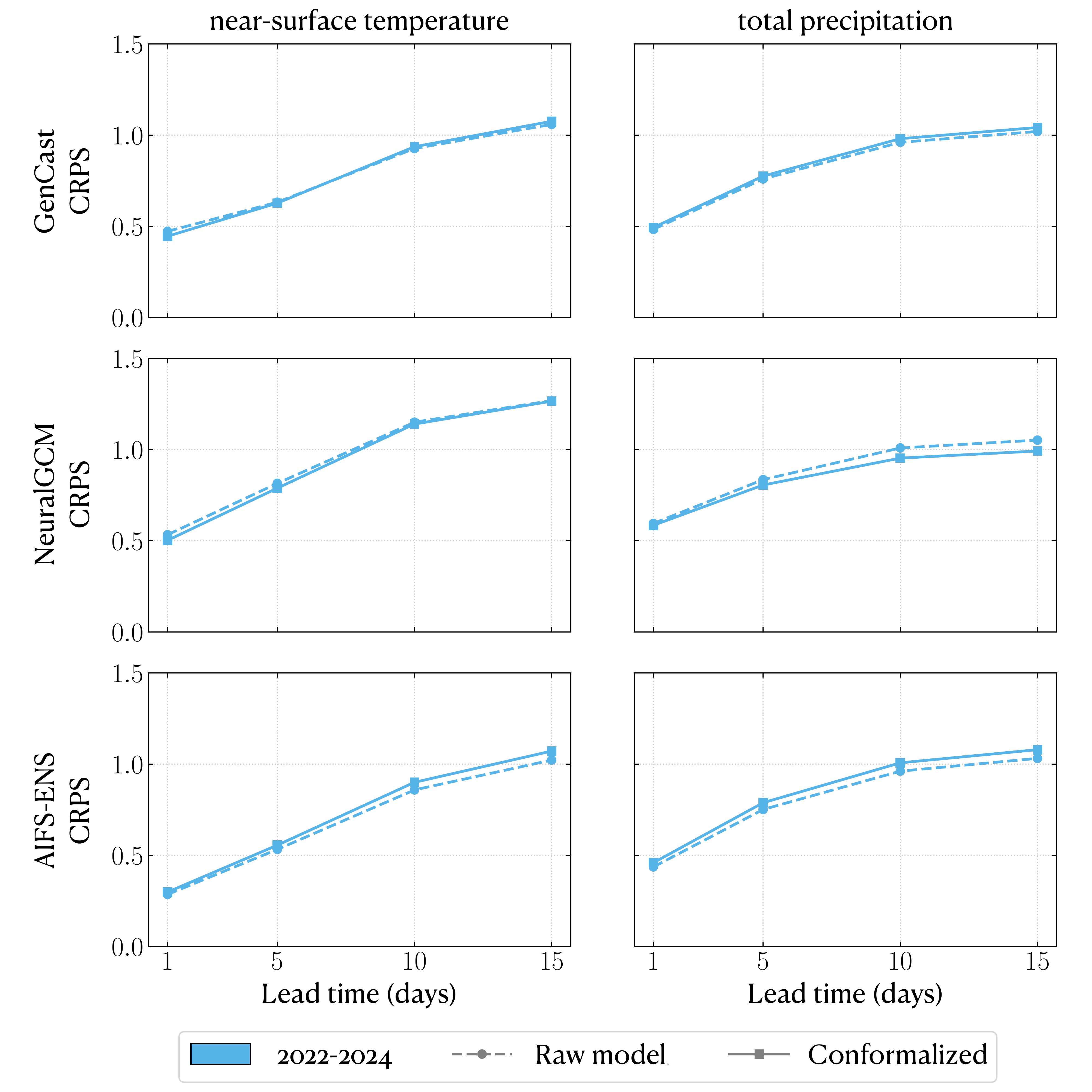}
\caption{The CRPS as a function of lead time, in units of the variable ($^\circ\textrm{C}$ for temperature, and $\textrm{mm}$ for precipitation). The performance of the raw model (dashed lines) is comparable to that of the conformalized model (solid lines).}
\label{fig:crps}
\end{figure}

\end{document}